\documentclass[aps,prl,twocolumn,preprintnumbers,superscriptaddress,nofootinbib]{revtex4-2}

\usepackage[utf8]{inputenc}
\usepackage{enumitem}

\usepackage{youngtab}

\usepackage[colorlinks=true,linkcolor=blue,citecolor=blue]{hyperref}%

\usepackage{float}
\usepackage{amsfonts}
\usepackage{amstext}
\usepackage{amsmath}
\usepackage{amsbsy}
\usepackage{graphicx}
\usepackage{amssymb}
\usepackage{amsthm}
\usepackage{bm}
\usepackage{comment}
\usepackage{epsfig}
\usepackage{latexsym}
\usepackage{cancel} 
\usepackage{color}
\usepackage{xcolor}
\usepackage{cleveref}

\DeclareSymbolFont{AMSa}{U}{msa}{m}{n}
\DeclareSymbolFont{AMSb}{U}{msb}{m}{n}
\DeclareMathSymbol{\fieldR}{\mathalpha}{AMSb}{"52}

\newcommand{\beq}{\begin{eqnarray}}
\newcommand{\eeq}{\end{eqnarray}}
\newcommand{\bea}{\begin{eqnarray}}
\newcommand{\eea}{\end{eqnarray}}
\newcommand{\be}{\begin{equation}}
\newcommand{\ee}{\end{equation}}
\newcommand{\bq}{\begin{equation}}
\newcommand{\eq}{\end{equation}}

\newcommand{\nn}{\nonumber}

\def\6{\partial}

\def\6{\partial}

\usepackage{color}
    \definecolor{darkgreen}{rgb}{0,0.5,0}
    \definecolor{darkblue}{rgb}{0,0,0.6}
    \definecolor{purple}{rgb}{0.4,.2,0.7}

\newcommand{\dm}[1]{#1}

\newcommand{\fig}[1]{Fig.~\ref{#1}}

\newcommand{\eqn}[1]{(\ref{#1})}

\newcommand{\h}{h_0}
\newcommand{\fre}{f_0}

\newcommand{\rhogw}{\rho_{\rm{GW}}}

\begin{document}

\title{Megahertz Gravitational Waves from Neutron Star Mergers}

\author{Diego Blas}
\affiliation{Institut de F\'\i sica d’Altes Energies (IFAE), The Barcelona Institute of Science and Technology, Campus UAB, 08193 Bellaterra (Barcelona), Spain}
\affiliation{Instituci\'o Catalana de Recerca i Estudis Avan\c cats (ICREA), Passeig Llu\'\i s Companys 23,  ES-08010, Barcelona, Spain.}

\author{Jorge Casalderrey-Solana}
\affiliation{Departament de F\'\i sica Qu\`antica i Astrof\'\i sica (FQA),  Universitat de Barcelona (UB), Mart\'\i\  i Franqu\`es, 1, 08028 Barcelona, Spain.}
\affiliation{Institut de Ci\`encies del Cosmos (ICCUB),  Universitat de Barcelona (UB), Mart\'\i\  i Franqu\`es, 1, 08028 Barcelona, Spain.}

\author{David Mateos}
\affiliation{Departament de F\'\i sica Qu\`antica i Astrof\'\i sica (FQA),  Universitat de Barcelona (UB), Mart\'\i\  i Franqu\`es, 1, 08028 Barcelona, Spain.}
\affiliation{Institut de Ci\`encies del Cosmos (ICCUB),  Universitat de Barcelona (UB), Mart\'\i\  i Franqu\`es, 1, 08028 Barcelona, Spain.}
\affiliation{Instituci\'o Catalana de Recerca i Estudis Avan\c cats (ICREA), Passeig Llu\'\i s Companys 23,  ES-08010, Barcelona, Spain.}

\author{Mikel~Sanchez-Garitaonandia}
\email[]{mikel.sanchez@polytechnique.edu}
\affiliation{Departament de F\'\i sica Qu\`antica i Astrof\'\i sica (FQA),  Universitat de Barcelona (UB), Mart\'\i\  i Franqu\`es, 1, 08028 Barcelona, Spain.}
\affiliation{Institut de Ci\`encies del Cosmos (ICCUB),  Universitat de Barcelona (UB), Mart\'\i\  i Franqu\`es, 1, 08028 Barcelona, Spain.}
\affiliation{CPHT, CNRS, \'Ecole polytechnique, Institut Polytechnique de Paris, 91120 Palaiseau, France}

\date{\today}

\begin{abstract}
Neutron star mergers provide a unique laboratory for the study of strong-field gravity coupled to quantum chromodynamics in extreme conditions. The frequencies and amplitudes of the resulting gravitational waves encode invaluable information about the merger. Simulations to date have shown that these frequencies lie in the kilohertz range. They have also shown that, if quantum chromodynamics possesses a first-order phase transition at high baryon density, then this is likely to be accessed during the merger dynamics. Here we show that this would result in the nucleation of superheated and/or supercompressed bubbles whose subsequent dynamics  would produce gravitational waves in the megahertz range. We estimate the amplitude of this signal and 
\dm{compare it to the sensitivity of planned future detectors}.
\end{abstract}

\maketitle

\section{Introduction}
\label{intro}

The revolutionary discovery of gravitational waves (GW) has opened two unprecedented windows into the Universe: one into the strong-field regime of gravity, and another one into the dynamics of quantum matter. These two aspects are often intertwined. This  is beautifully illustrated by neutron star (NS) mergers, in which quarks and gluons described by quantum chromodynamics (QCD) interact in the strong gravitational field sourced by the quarks and gluons themselves. 

\begin{figure}[t]
   \begin{center}
   	\includegraphics[width=0.45\textwidth]{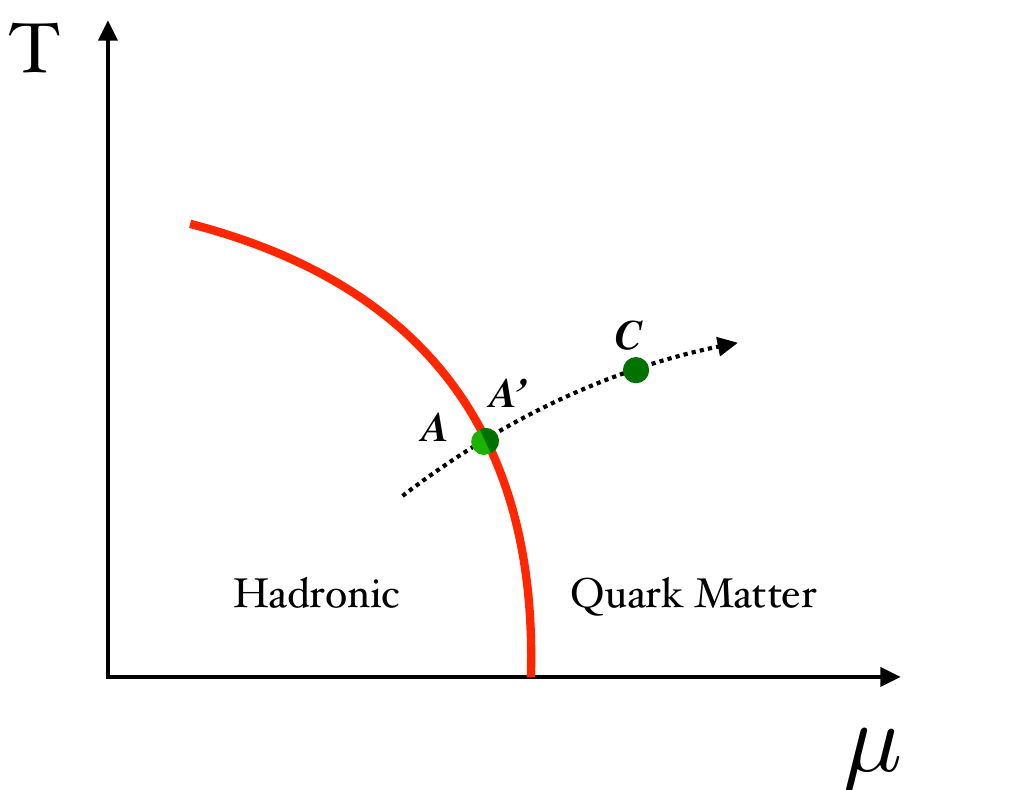} 
   \end{center}
   \caption{\small A possible phase transition in QCD, indicated by the solid red curve.  $T$ and $\mu$ are the temperature and the baryon chemical potential, respectively.  The dotted black curve shows a possible evolution of a region of a NS merger as this region is heated and/or compressed.  The points dubbed $A$, $A'$ and $C$ correspond to the states shown in \fig{meta}.}
\label{FOPT}
\end{figure}

Most of the merger dynamics can be accurately described by numerical simulations of magneto-hydrodynamics coupled to general relativity (for a review, see e.g.~\cite{Baiotti:2016qnr}). One robust conclusion of these simulations is that the characteristic lifetime of the system is of order of several, possibly 10, milliseconds (ms). These simulations take as an input the properties of QCD, in particular its Equation of State (EoS). At zero baryon density the EoS can be reliably determined via the lattice formulation of QCD. In contrast, the famous sign problem \cite{deForcrand:2009zkb} renders this tool inapplicable at the baryon densities of interest for NS physics. In particular, it is unknown whether QCD possesses first-order phase transitions (FOPT) in this region. Nevertheless, a variety of arguments suggest that at least two FOPT may be present \cite{Stephanov:2004xs,Alford:2007xm,Fukushima:2010bq,Guenther:2022wcr}, one of which is sketched in \fig{FOPT}. One is the transition from hadronic matter to deconfined, quark matter. The other is the transition from a non-superconducting 

to a color-superconducting phase.  We emphasize that, while these transitions are well motivated, at the theoretical level their existence cannot be rigorously established at present. NS mergers have the potential to establish this experimentally, possibly in combination with heavy ion collision experiments. To realise this potential, a theoretical understanding of the imprints of a phase transition on the GW spectrum is indispensable. 

Numerical simulations of NS mergers based on EoS with a hadronic-quark matter phase transition include \cite{Most:2018eaw,Most:2019onn,Ecker:2019xrw,Prakash:2021wpz,Weih:2019xvw,Tootle:2022pvd, Fujimoto:2022xhv}. These studies have shown that the dynamics of the merger results in the formation of regions in which the matter is sufficiently heated and/or compressed  so that the thermodynamically preferred phase is the quark-matter phase. We will generically refer to these regions as ``Hot or Compressed Spots'', or HoCS for short 
The characteristic size of these HoCS is \cite{Tootle:2022pvd}
\be
\label{L}
L \simeq 5 \, \mbox{km} \,.
\ee

 To our knowledge, no simulation based on an EoS that includes a transition to a color-superconducting phase has been performed. Nevertheless, existing simulations  have shown that the merger leads to the formation of cold quark-matter regions in which the baryon density can be ten times larger than the nuclear saturation density.
This makes it conceivable that color-superconducting matter may be formed in NS mergers.

Numerical simulations have shown that the \emph{presence} of HoCS affects the overall merger dynamics, mainly because of a softening of the EoS, and leads to detectable modifications of the GW spectrum in the usual kilohertz (kHz) frequency range. One may think of this as a ``macroscopic'' consequence of the phase transition. However, to the best of our knowledge, the \emph{phase transition dynamics} in   HoCS has not been previously considered. The purpose of this paper is to show that this dynamics leads to a characteristic signal in the megahertz (MHz) range. One may think of this as a ``microscopic'' consequence of the phase transition. We will see that this conclusion is extremely robust since it only depends on generic properties of a FOPT. Therefore  we do not need to commit ourselves to a particular type of FOPT. The only assumptions we need to make are that some FOPT is present in QCD and that this is accessed by the merger dynamics. 

The key idea is that the millisecond characteristic time scale for the evolution of the merger,
\be
\label{char}
\tau\simeq 1 \textrm{ms}\,,
\ee 
is much longer than the characteristic nuclear time scale, \mbox{1 fm $\simeq$ $10^{-23}$ s}. This huge separation of scales means that, from the viewpoint of QCD processes, the merger evolution is adiabatic to an extremely good approximation. In turn, this implies that the HoCS are initially born as carefully prepared  metastable regions of  superheated and/or supercompressed matter. Once the superheating or supercompression are large enough, namely once the HoCS is sufficiently deep into the metastable branch, bubbles of the stable phase begin to nucleate. The point where this happens is labeled ``$B$'' in  \fig{meta}. These bubbles then expand to a macroscopic size and collide, leaving behind long-lived sound waves propagating on top of the stable phase, as illustrated in \fig{sound_waves}.  As we will explain below, this dynamics gives rise to a characteristic GW spectrum whose peak  frequency is in the MHz range. 

As usual, we work with units such that $\hbar=c=1$. $G$ will be Newton's constant, related to the reduced Planck mass through 
\be
(8\pi G)^{-1/2} = M_p \simeq  2.4 \times 10^{18} \, \mbox{GeV} \,.
\ee 

\section{Frequency}
\label{freq}

We begin by noting that the dynamics of a FOPT in a NS merger is very similar to that of a cosmological FOPT (for a review, see e.g.~\cite{Hindmarsh:2020hop}), except for the fact that in the cosmological case the metastable phase is supercooled. In  this case there is a compelling physical picture that emerges from the large body of work developed over several decades. Here we will take advantage of this picture to give an order-of-magnitude estimate of the GW signal in a NS merger.

In the cosmological case the separation of scales is provided by the fact that the expansion rate of the Universe, $H^{-1}$, with $H$ the Hubble rate, is much longer than the microscopic time scale given by the inverse of the local temperature, $T^{-1}$. As the Universe expands and cools down, it eventually enters the metastable phase. At some point bubbles of the stable phase begin to nucleate.These bubbles then grow and collide leaving behind a superposition of long-lived sound waves propagating on the stable phase. At sufficiently late times turbulence may also develop. Although each of these processes contributes to the GW spectrum, for many models of the transition it has been shown that the dominant contribution comes from collisions of sound waves with one another \cite{Hindmarsh:2013xza}. As a first approximation, we will assume that the same may be true in a NS merger. We will now determine the peak frequency of the GWs, and in the next section we will estimate their characteristic strain. 
\begin{figure}
   \begin{center}
   	\includegraphics[width=0.45\textwidth]{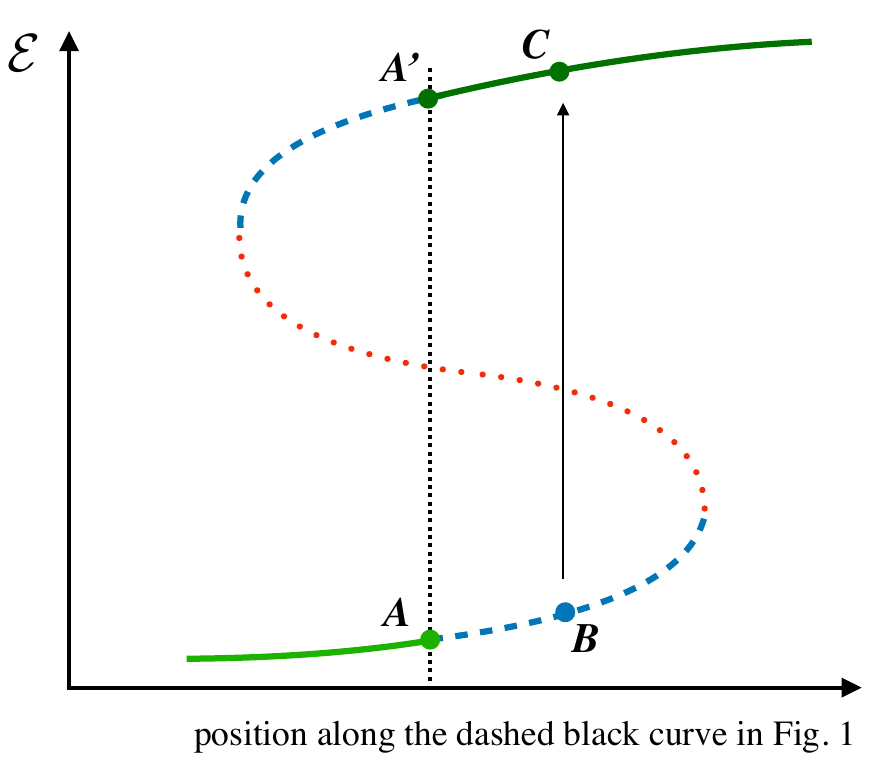} 
   \end{center}
    \caption{\small Energy density as a function of the position along the dashed black curve in \fig{FOPT}. Both $T$ and $\mu$ increase from left to right. The dotted black vertical line indicates the location of the phase transition, determined by the condition that the states $A$ and $A'$ have the same free energy density. Stable, metastable and unstable states are shown as solid green, dashed blue and doted red respectively. As a region is heated/compressed it enters the lower metastable branch.  At $B$  the nucleation probability has been sufficiently enhanced, and bubbles of the preferred state $C$ on the upper stable branch are quickly nucleated.} 
\label{meta}
\end{figure}

\begin{figure*}[t!!!]
   \begin{center}
   	\includegraphics[width=0.98\textwidth]{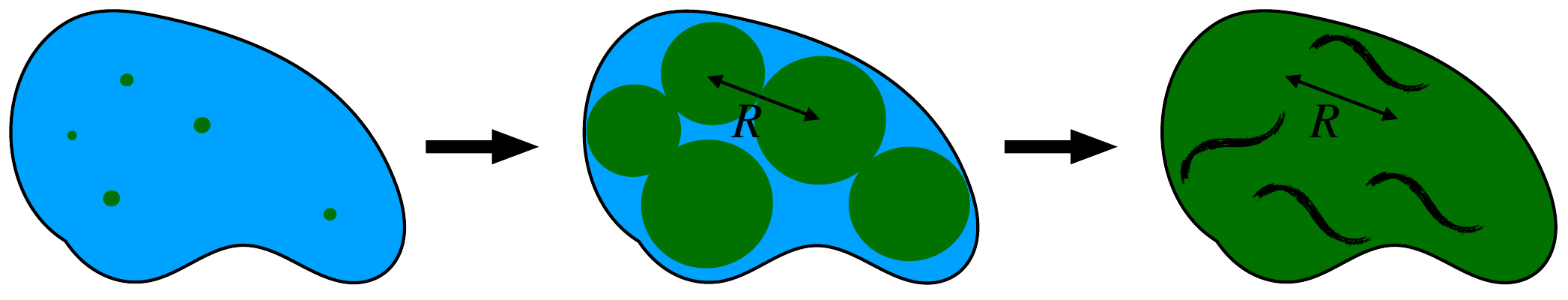} 
   \end{center}
    \caption{\small FOPT dynamics inside a HoCS. Once a HoCS enters the metastable region, it takes a time of order $\tau \simeq 1$ ms for the first few bubbles to nucleate inside the metastable phase (left). Bubbles then grow to a macroscopic size $R$ and collide in a time $\beta^{-1} \simeq 6\,  \mu$s (center). They leave behind long-lived sound waves (with lifetime $\tau \simeq 1$ ms) of characteristic size $R$ propagating on top of the stable phase (right).}
\label{sound_waves}
\end{figure*}

Let $R$ be the mean bubble-centre separation at the end of the phase transition, as in \fig{sound_waves}. 
This scale is inherited by the sound waves left behind after the bubbles collide, and it is further imprinted on the GWs produced by the subsequent fluid dynamics. The peak GW wavelength, is therefore $\lambda \simeq R$ \cite{Hindmarsh:2020hop,Hindmarsh:2015qta}.

Nucleated bubbles grow with a constant bubble wall velocity $v_w$, which is usually assumed to be a fraction of the speed of light --- see e.g.~\cite{Cutting:2019zws}. We therefore take the ballpark value 
\be
\label{vv}
v_w \simeq 0.1 \,.
\ee
This estimate is consistent with holographic simulations \cite{Bea:2024bls}. Changing it has a simple effect on the peak GW frequency --- see~(\ref{fre}). 

Let $\beta^{-1}$ be the duration of the phase transition, whose precise definition is in~\eqn{betabeta}. Intuitively it can be understood as the time from the moment that the first few bubbles are nucleated to the moment when all the bubbles have collided and disappeared, leaving behind sound waves on top of the stable phase, as illustrated in  \fig{sound_waves}.  The duration of the phase transition is related to the mean bubble separation through \citep{Hindmarsh:2020hop}
\be
\label{RR}
R = \left( 8\pi \right)^{1/3} \frac{v_w}{\beta} \,.
\ee
Consequently, in order to determine the GW wavelength, we must estimate the duration of the phase transition, for which we need to recall how the phase transition takes place dynamically.

Suppose that some region of the NS follows the trajectory indicated by the dashed black curve in \fig{FOPT}(left). The energy density of the available phases of the system as $T$ and $\mu$ vary along this curve is shown in \fig{meta}, where the multivaluedness is a hallmark of a FOPT. The different colors in  \fig{meta} indicate stable, metastable and unstable states. Note that only the stable states are shown in the phase diagram of \fig{FOPT}(left), whose hadronic and quark-matter  phases correspond to the lower and upper stable branches of \fig{meta}, respectively.  Points $A$ and $A'$ are  reached as the phase transition curve is approached along the hadronic and the quark-matter  phases, respectively. The qualitative details of \fig{meta} do not depend on the specific trajectory crossing the FOPT line.

As the region in the NS is heated and/or compressed along the trajectory of \fig{FOPT}(left), the corresponding state moves along the lower stable branch in \fig{meta} until it eventually enters the lower metastable branch. Once this happens, bubbles of the stable phase can begin to nucleate. On general grounds, the probability per unit volume and per unit time for this to happen takes the form
\be
\label{prob}
\frac{dP}{dt \, d^3 x } = \Lambda^4 \, e^{-S[\Lambda]}.
\ee
The exact value of the $\Lambda^4$ prefactor  depends on the  specific point on the metastable branch, but its scale is set by the characteristic values of the energy and pressure in NS matter. At zero temperature these are of order $\mathcal{E} \simeq \mathcal{P} \simeq 0.5 \, \mbox{GeV}/\mbox{fm}^3\,$\cite{Gorda:2022lsk}.

Since a non-zero temperature will increase these values, we take as a  plausible range 
\be
\label{Lambda}
0.1\, \mbox{GeV}/\mbox{fm}^3 \lesssim \Lambda^4 \lesssim  1\, \mbox{GeV}/\mbox{fm}^3 
\ee
and adopt as a reference value the intermediate point $\Lambda^4\simeq 0.5$ GeV/fm$^3$. 

$S$ is the action of the critical bubble, defined as the bubble for which the inward-pointing force due to the surface tension is exactly balanced by the outward-pointing force due to the pressure difference between the inside and the outside of the bubble. The action of this bubble controls the  nucleation probability in the semiclassical approximation, hence the exponential dependence in \eqn{prob}. 
$S$ is infinite at the point $A$ where the metastable phase meets the stable phase, and it decreases until it reaches zero at the turning point where the metastable phase meets the unstable phase. At an arbitrary point on the metastable branch the value of $S$ depends on the corresponding values of the temperature and the chemical potential. The exponential enhancement of the nucleation probability with time implies that the transition will happen rapidly around a time $t_f$ (point B in \fig{meta}). By performing a Taylor expansion around $t_f$ we obtain \cite{SM,Enqvist:1991xw,Hindmarsh:2020hop}
\be
\label{betabeta}
\beta^{-1}  = \left(\frac{dS}{dt}\right)^{-1} \simeq \frac{\tau}{S} \simeq\frac{\tau}{\log\left(8\pi v_w^3\tau^4\Lambda^4\right)}\,,
\ee
where we have assumed that $\Lambda dS/d\Lambda \sim S$.

Plugging \eqn{betabeta} into \eqn{RR} we arrive at the following expression for the peak frequency $\fre=\lambda^{-1}$:
\begin{widetext}
\be 
\label{fre}
\fre \simeq \frac{0.1}{v_w}\frac{1\, \text{ms}}{\tau}\left(0.62 + 0.0034\log\left[\left(\frac{v_w}{0.1}\right)^3\left(\frac{\tau}{1\, \text{ms}}\right)^4\frac{\Lambda^4}{0.5\, \text{GeV}/\text{fm}^3}\right]\right)\text{MHz}\,.
\ee
\end{widetext}
Using the reference values given above leads to 
\be
R \simeq 477\,\textrm{m}\,,\,\,\, \beta^{-1}\simeq 5.43\, \mu \textrm{s}\,,\,\,\, 
\fre \simeq 0.6\, \textrm{MHz} \,.
\ee 
These values are fairly robust. Up to small logarithmic corrections, they depend in a simple way on $\tau$ and $v_w$. The former is well constrained by simulations. The main uncertainty comes from  $v_w$. Allowing this to increase or decrease by an order of magnitude around the ballpark value \eqn{vv} results in a peak frequency roughly between 
\be
\label{ff}
100 \, \textrm{kHz} \lesssim \fre \lesssim 10\,  \textrm{MHz} \,.  
\ee

To recap, when a region enters the metastable region, bubbles start nucleating after a time of order a milisecond, see \fig{sound_waves}-left. The typical number of nucleated bubbles in each regions is of order $N_{bubble} =  L^3/R^3  \simeq 1150$. Once nucleation stars, it takes few microseconds for the FOPT to complete, namely to go from left to right in \fig{sound_waves}. The lifetime of the sound waves left behind after the transition in \fig{sound_waves}-right is of the order of miliseconds too. They represent the dominant contribution to GW emission, which is peaked at a frequency of order a MHz.

\section{Characteristic strain}
\label{amp}

We follow the physical picture developed in the cosmological case and assume that the dominant GW production mechanism is due to the long-lived sound waves left behind after all the bubbles have collided. The produced GW energy density and its relation to the characteristic strain (see e.g.~\cite{Moore:2014lga}), can be estimated as
\be
\label{rho}
\rhogw = \frac{2\pi^2}{8\pi G} \, \frac{\h^2}{\lambda^2} \simeq ( 8\pi G)  \, v_f^4 \, (\mathcal{E}+\mathcal{P})^2 \, \Delta t \,R \,\bar{\Omega}_{\rm{GW}} \,.
\ee
A detailed derivation of this formula can be found in e.g.~\cite{Hindmarsh:2015qta}. We will not reproduce the derivation here, but we will provide intuition for the origin of each term.

$\mathcal{E}$, $\mathcal{P}$ and $v_f^2$ are the energy density, the pressure and the root mean velocity of the fluid in the HoCS left behind by the bubble collisions, as illustrated in \fig{sound_waves}(right). We take $\mathcal{E}\simeq \mathcal{P} \simeq \Lambda^4$ with $\Lambda^4$ in the range \eqn{Lambda}.

GWs are metric perturbations sourced by the transverse-traceless part of the energy-momentum tensor, which for sound waves on a fluid $h _{ij}\sim 8\pi G T^{TT}_{ij} \sim 8\pi G(\mathcal{E}+\mathcal{P})v_i v_j$, leading to the $8\pi G v_f^4 \, (\mathcal{E}+\mathcal{P})^2$ factor in \eqn{rho}.

$\Delta t$ is the effective time duration of the source for the purpose of determining $\rhogw$. In the cosmological case  \mbox{$\Delta t \simeq H^{-1}$}. In the NS case, however, GW contributing to the growth of $\rhogw$ leave the HoCS after a light-crossing time of the region, shorter than the characteristic millisecond time of the merger,
\be
\label{deltat}
\Delta t \simeq L \simeq 1.7 \times 10^{-2} \, \mbox{ms} \,,
\ee
where we have used \eqn{L}.

$R$ is a necessary characteristic length scale in the fluid flow on dimensional grounds, which has been seen to coincide with the mean separation between the bubbles that produces the fluid flow in the first place.

Finally, $\bar{\Omega}_{\rm{GW}}$ is a dimensionless factor quantifying the efficiency with which shear stress in the fluid is converted to GWs. Although  its precise value depends on the details of the flow, simulations show that its ballpark value, $\bar{\Omega}_{\rm{GW}}\sim 10^{-2}$, is rather insensitive to these details.

Extracting the strain from \eqn{rho}, and taking into account that the GW amplitude decreases with the distance $d$ to the source, we arrive to the observed characteristic strain,
\begin{widetext}
\be
  \h^{\rm{obs}} \,\,\, = \,\, \,\h\, \frac{L}{d} \,\,\, \simeq \,\, \,6.2\times 10^{-24} \, v_f^2 \, \left(\frac{\Lambda^4}{0.5\ \text{GeV}/\text{fm}^3}\right)\left(\frac{L}{5\ \text{km}}\right)^{3/2}\left(\frac{0.6\ \text{MHz}}{f}\right)^{3/2}\left(\frac{100\ \text{Mpc}}{d}\right)\,.
  \label{peakamp}
\ee
\end{widetext}
The fluid velocity depends on details of the phase transition such as the strength of the transition, the bubble wall velocity, etc., but a representative range can be taken to be \cite{Cutting:2019zws}
\be
\label{vrange}
0.01 \lesssim v_f \lesssim 0.3 \,,
\ee
with higher values favoured for strong phase transitions, as might be expected for QCD. 

\dm{The strain estimated above corresponds to that at the peak. Away from the peak frequency, the strain is expected to decrease following a power law  with width $\delta f \sim \fre$ \cite{Hindmarsh:2019phv}.}

\section{Detector sensitivities}

Our signal lies in a MHz band that is highly active in proposed concepts for gravitational-wave detection \cite{Aggarwal:2020olq,Aggarwal:2025noe}.
According to \cite{Aggarwal:2025noe}, detectors operating in this band—such as magnetic Weber bars \cite{Domcke:2024mfu}—could feasibly achieve a characteristic noise-equivalent strain (PSD) of order \mbox{$S_n(f)^{1/2} \sim 2\times 10^{-22}\, \rm Hz^{-1/2}$}.
See also \cite{Berlin:2023grv,Aggarwal:2020umq} for additional concepts that may achieve comparable sensitivities. In this section  we will compare the sensitivities of these detectors with \eqn{peakamp} evaluated with the upper values in the ranges \eqn{Lambda} and \eqn{vrange}. 

For a transient source of duration $\tau$ with a broad spectrum of width $\delta f$ centered at $f_0$ and characteristic strain $h_0$, the expected sensitivity to $h_0$ from a broadband search is
\cite{Maggiore:2007ulw,Anderson:2000yy,Flanagan:1997sx}
\begin{equation}
\label{broad}
\begin{split}
        &  \h^{\rm{broad}} \gtrsim \frac{1}{\tau}\left(\frac{S_n(f_0)}{\pi f_{0}}\right)^{1 / 2}{\cal N}^{1/2}\\
       &    \sim 
       10^{-21}\left(\frac{\rm m s}{\tau}\right)
       \left( \frac{S_n(f_0)^{1/2}}{10^{-22}\,\rm Hz^{-1/2}}\right)
       \left(\frac{\rm 0.6 \, MHz}{f_0}\right)^{1/2}\frac{{\cal N}^{1/2}}{10}\, ,
\end{split}
\end{equation}
where
\begin{equation}
{\cal N}\sim 1.6\times 10^2  \left(\frac{\tau}{\rm ms}\right)\left(\frac{\delta f}{0.6\, \rm MHz}\right),
\end{equation}
denotes the number of bins to be fitted. Since $\delta f \sim f_0$, the resulting sensitivity is approximately flat across the entire frequency band of interest.

For a resonant search, the prospects for magnetic Weber bars discussed in \cite{Domcke:2024mfu} assume a quality factor of $Q \sim 2\times 10^7$, corresponding to a very narrow resonance width $\Delta f \sim f_0/Q$. In this case \cite{Maggiore:2007ulw}
\begin{equation}
\label{resonant}
    \begin{split}
        &   \h^{\rm{resonant}} \gtrsim \frac{1}{\tau}\left(\frac{S_n(f_0)}{\pi \Delta f}\right)^{1 / 2}\rm 
        \times SNR\\
        &\sim 10^{-21}\left(\frac{\rm m s}{\tau}\right)
        \left(\frac{\rm 0.6 \, MHz}{f_0}\right)
        \left(\frac{Q}{2\times 10^7}\right)^{1/2},
       \end{split}
\end{equation}
where we used that on resonance \cite{Domcke:2024mfu}
\begin{equation}
S_n(f_0)^{1/2}\sim {4\times 10^{-25}
\,\rm Hz^{-1/2}} \left( 
\frac{0.6{\rm \,MHz}}{f_0} \right)^{1/2} \,,  
\end{equation}
and we have taken ${\rm SNR}\sim 1$.

\begin{figure}[!tb]
\includegraphics[width=.4\textwidth]{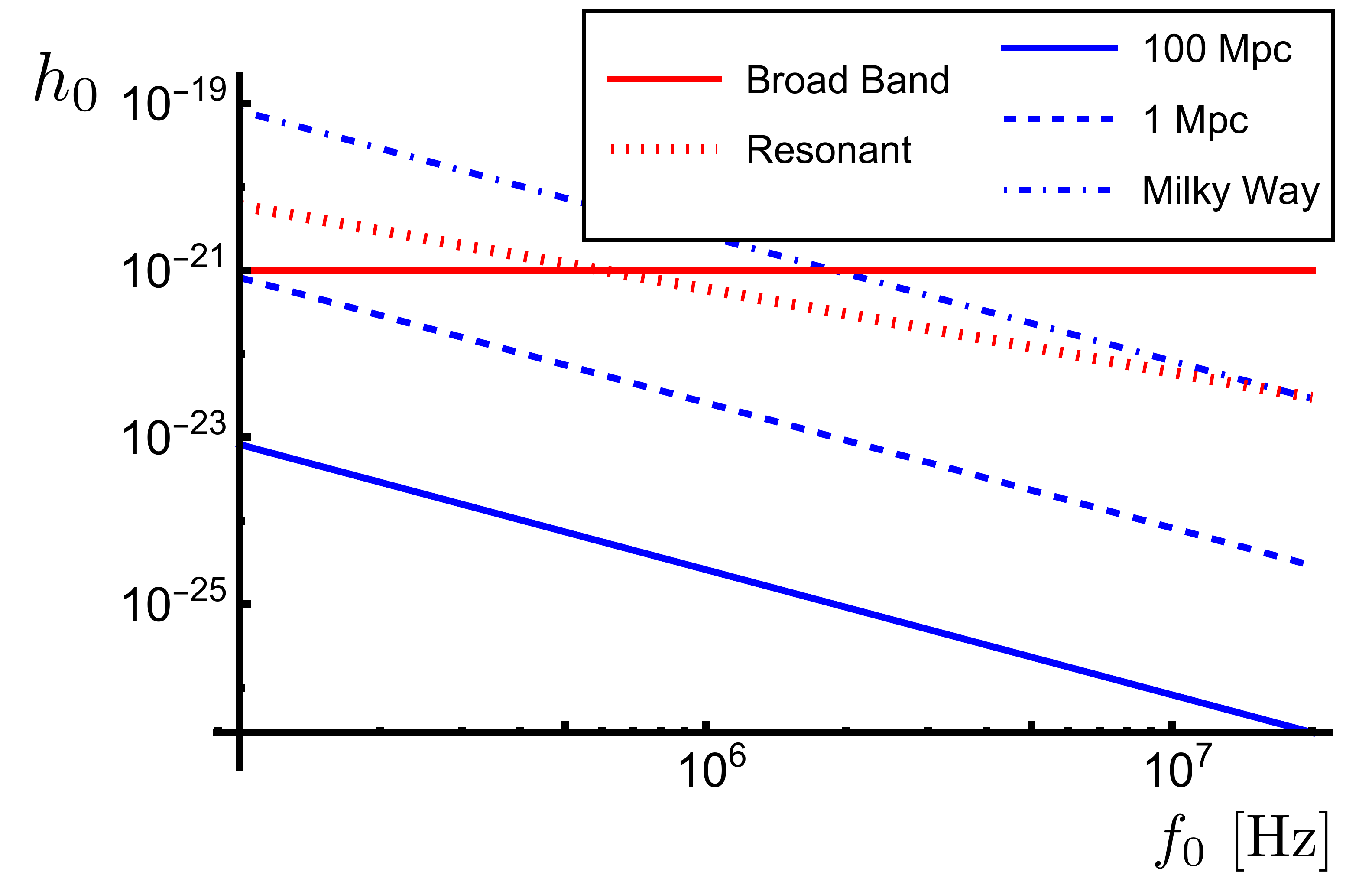}
\vspace{-2ex}
\caption{
Sensitivity of the proposal in \cite{Domcke:2024mfu} (broadband and resonant) compared with the expected signal from an NS–NS collision at various distances from Earth.}
\label{fig:reach}
\vspace{-2ex}
\end{figure}

Fig.~\ref{fig:reach} shows a comparison between the predicted signal \eqn{peakamp} and the sensitivities \eqn{broad} and \eqn{resonant}. We will come back to this plot below.

\section{Discussion}
\label{disc}

Numerical simulations suggest that a QCD FOPT is accessed during NS merger dynamics. Due to the separation between the merger dynamics characteristic time scale and the nuclear one, the evolution can be considered adiabatic from the QCD viewpoint. This results in the formation of metastable regions of superheated and/or supercompressed matter in which the phase transition takes place via the formation and subsequent expansion and collision of bubbles. 

This physics is analogous to that of a cosmological FOPT, except that in the latter, the metastable region is supercooled. Exploiting this analogy we provided a rough estimate of the features of the resulting GW spectrum. We found that the duration of the signal is of order of milliseconds and the peak frequency is in the MHz range. To date, this is the only astrophysical source of GWs in this frequency range that is based exclusively on Standard Model physics. Analogously to the cosmological case, we expect the precise GW power spectrum to be a broken power law centred around this frequency.

Since bubbles are initially microscopic, for almost all of their evolution they behave as in flat space. Thus, the main effect of the spacetime curvature in the NS is simply to redshift the GW frequency. This does not change the order of magnitude of our estimate --- for reference, the frequency of a GW emitted from the surface of a $2M_\odot$ NS would be reduced by 36\%. 

The characteristic strain that would be observed coming from a single HoCS is given by \eqn{peakamp}. Simulations show that several spots are typically formed, and that they can transition back to the original phase via the nucleation and collision of supercooled bubbles, leading to additional emission of GWs in a similar frequency range. We expect these effects to enhance the signal that we have estimated. A further enhancement could come from the possible appearance of turbulence triggered by the dynamics of the sound waves. 

The detection prospects are summarized in \fig{fig:reach}. Broadband searches are more effective at frequencies below $0.6$ MHz, while resonant searches perform better above this frequency. The sensitivity curves in \fig{fig:reach} could be further improved by employing more sophisticated burst-detection methods (see e.g. \cite{Drago:2020kic,Sutton:2009gi,Cornish:2014kda}), and by using multiple detectors $N_{\rm det}$, which can enhance the sensitivity by a factor of $\sqrt{N_{\rm det}}$. Moreover, because the merger time is known from the preceding inspiral, coincident methods can further enhance detection efficiency. Nevertheless, detector glitches and other instrumental limitations may reduce this improvement.

Although the  prospects above fall short of guaranteeing a near-term detection, the growing interest in high-frequency GW searches holds the potential for groundbreaking results. Improving current approaches by up to three orders of magnitude appears plausible, especially given the possibility of detecting a signal that would open a new window for probing hot and dense matter.

It would be interesting to study other effects of the phase transition such as the possible emission of neutrinos or the gravitational wave memory. 

\section*{Acknowledgments}
\begin{acknowledgments}
We are grateful to Jordi Miralda for collaboration at the initial stages of this paper, and to Krishna Rajagopal for comments on the draft. We thank Mark Alford, Yago Bea, Neil Cornish, Sebastian Ellis, Roberto Emparan, Alexander Haber, Luciano Rezzolla, Andreas Schmitt, Alexandre Serantes, Leo Stein, and specially Andreas Ringwald and Carlos Tamarit, for useful discussions. 
We are supported by the ``Unit of Excellence MdM 2020-2023'' award to the Institute of Cosmos Sciences (CEX2019-000918-M) and by grants PID2019-105614GB-C21, PID2019-105614GB-C22, PID2022-136224NB-C21, PID2022-136224NB-C22, and 2021-SGR-872. This work is supported by the European Research Council (ERC) under the European Union’s Horizon 2023 research and innovation programme (grant agreement No  101141909).
MSG is also supported by the European Research Council (ERC) under the European Union’s
Horizon 2020 research and innovation program (grant
agreement No758759).

This publication is part of the grant PID2023-146686NB-C31 funded by MICIU/AEI/10.13039/501100011033/ and by FEDER, UE.
IFAE is partially funded by the CERCA program of the Generalitat de Catalunya.

This work is supported by ERC grant (GravNet, ERC-2024-SyG 101167211, DOI: 10.3030/101167211). Funded by the European Union. Views and opinions expressed are however those of the author(s) only and do not necessarily reflect those of the European Union or the European Research Council Executive Agency. Neither the European Union nor the granting authority can be held responsible for them. 

D.B. acknowledges the support from the European Research Area (ERA) via the UNDARK project of the Widening participation and spreading excellence programme (project number 101159929).
\end{acknowledgments}

\bibliography{refsHolographicBubbles-2}

\onecolumngrid
\appendix
\section{Supplemental Material}
In this supplemental material we display in detail the derivation of the expressions for the duration of the phase transition, $\beta^{-1}$, and the mean bubble separation, $R$, in the letter. The calculation is an adaptation of the same computation performed in the context of cosmological phase transitions and can be found in \cite{Enqvist:1991xw, Hindmarsh:2020hop}, for example.

\section{Duration of the phase transition}
Let $t_c$ be the time when the system enters the metastable phase. At a later time $t>t_c$, the fraction of the HoCS volume still in the metastable phase is 
\cite{Guth:1981uk, Enqvist:1991xw}
\be
\label{qqq}
q(t) = e^{-I(t)}  \,,
\ee
with
\be
\label{III}
I(t) = \int_{t_c}^t \, dt' \, \frac{4}{3} \pi v_w^3 (t-t')^3 \, 
\Lambda^4 e^{-S(t')}  \,.
\ee
Eqn.~\eqn{qqq} is easily interpreted. The integrand 
in \eqn{III} is the volume of a bubble nucleated at a time $t_c< t'<t$ multiplied by the bubble nucleation probability at that time. At early times $I(t)\ll 1$ and  
\be
q(t) \simeq 1- I(t) \,.
\ee
This is the expected expression  if bubble overlaps are ignored. The exponentiation in \eqn{qqq} accounts for these overlaps. 

As time progresses beyond $t_c$ the system penetrates deeper into the metastable branch and the action $S$ decreases. This implies that the integrand in \eqn{III} is dominated by times $t'$ near $t$. We can therefore  Taylor-expand the action to linear order in $t-t'$ to obtain
\be
\label{hap}
q(t) \simeq \exp \left[ - \frac{4}{3} \pi v_w^3 \, \Lambda^4 e^{-S(t)} 
\, 6\beta^{-4} \right] \,,
\ee
where
\begin{equation}
\label{beta}
\beta = -\frac{dS}{dt}\,.
\end{equation}
The phase transition takes place at the time $t_f$ where the exponent in \eqn{hap} becomes unity, namely when 
\be
\label{unity}
8 \pi v_w^3 \, \Lambda^4 e^{-S(t_f)} 
\, \beta(t_f)^{-4} = 1 \,.
\ee
At this time a fraction $1/e \simeq 37\%$ of the volume remains in the metastable phase. Expanding \eqn{hap} around $t_f$ we see that $q(t)$ takes the very simple form
\be
\label{und}
q(t) \simeq \exp \left[ -e^{ \beta (t-t_f) } \right] \,.
\ee
In this and in subsequent equations, $S$ and $\beta$ are understood to be evaluated at $t=t_f$. Eqn.~\eqn{und}  justifies the interpretation of $\beta^{-1}$  as the  duration time of the transition. This time can be determined from the condition \eqn{unity}. To do so, we first use the chain rule to rewrite \eqn{beta} as 
\be
\label{dim}
\beta = \left( - \Lambda \frac{dS}{d\Lambda} \right) 
\left(\frac{1}{\Lambda}\frac{d\Lambda}{dt} \right)\,.
\ee
The first factor on the right-hand side is purely microscopic since it measures the variation of the microscopic action with the microscopic scale. The second factor is simply the expansion rate of the system 
\begin{equation}
\label{exp}
\frac{1}{\tau} = \frac{1}{\Lambda}\frac{d\Lambda}{dt} \,.
\end{equation}
The time $\tau$ is the characteristic evolution time of the system which, 
based on NS merger simulations, is of order 
\be
\label{tau}
\tau \simeq 1\, \mbox{ms} \,.
\ee
Generically, we expect the dimensionless derivative of $S$ in the first factor on the right-hand side of \eqn{dim} to be of the same order as $S$ itself. Therefore 
\be
\label{bb}
\beta \tau \simeq S \,.
\ee
Substituting in \eqn{unity} we then have
\be
S^4 \, e^{S} \simeq 8\pi \, v_w^3 \tau^4 \Lambda^4 \,,
\ee
whose solution, approximating the left hand side by the exponential term, is

\be
S \simeq \log\left(8\pi v_w^3\tau^4\Lambda^4\right).
\ee
As a reference, the value for the action once we plug in all the reference values we are considering ($v_w\sim 0.1$, $\Lambda^4\sim 0.5\mathrm{GeV}/\text{fm}^3$, $\tau \sim 1$ms) is $S \simeq 184$, which implies that $\beta^{-1} \simeq 5.43\mu$s,  as presented in the main text.

We can now determine the mean bubble center separation. The number of bubbles per unit volume at a time $t>t_c$ is the integral until that time of the probability of nucleating bubbles in the metastable phase, times the available volume of metastable phase. Thus at a time $t>t_c$ this is given by 
\be
n_{bubble}(t) = \int_{t_c}^t dt' \, q(t') \, \frac{dP}{dt' \, d^3 x } \,.
\ee
Using the Taylor expansions around $t_f$ introduced above, this becomes

\bea
n_{bubble}(t) &=& \Lambda^4 e^{-S(t_f)} \int_{t_c}^t dt' \, \exp \left[ -e^{ \beta (t-t_f) } \right] 
\, e^{ \beta (t-t_f) } 
 = -  \Lambda^4 e^{-S(t_f)} \beta^{-1} 
\int_{t_c}^t dt' \, \frac{d}{dt'} \exp \left[ -e^{ \beta (t-t_f) } \right] \nn\\[2mm]
&=& \Lambda^4 e^{-S(t_f)} \beta^{-1} \Big[ 1 -q(t) \Big] \,.
\eea

At late times $q(t\to \infty) \to 0$ and, using \eqn{unity}, we arrive at the final density of bubbles
\be
n_{bubble} = \frac{\beta^3}{8\pi v_w^3}\, ,
\ee
from which the mean bubble size at the time of collision follows,

\be
R = (n_{bubble})^{-1/3} = (8\pi)^{1/3}\frac{v_w}{\beta}.
\ee

\end{document}